\documentclass[a4paper,10pt,twoside]{cpc-hepnp}

\usepackage{multicol}
\usepackage{graphicx}
\usepackage{booktabs}
\usepackage{amssymb,bm,mathrsfs,bbm,amscd}
\usepackage[tbtags]{amsmath}
\usepackage{lastpage}

\begin{document}

\fancyhead[co]{\footnotesize B. Julia-Diaz: Single photo and electroproduction of pions at EBAC@JLAB }

\footnotetext[0]{Received 20 June 2009}

\title{Single photo and electroproduction of pions at EBAC@JLAB}

\author{%
      B. Julia-Diaz$^{1,2)}$
}
\maketitle

\address{%
1~(Department d'Estructura i Constituents de la Mat\`{e}ria
and Institut de Ci\`{e}ncies del Cosmos,
Universitat de Barcelona, E--08028 Barcelona, Spain   )\\
2~(Excited Baryon Analysis Center (EBAC), Thomas Jefferson National
Accelerator Facility, Newport News, VA 23606, USA)\\
}

\begin{abstract}

Within the Excited Baryon Analysis Center we have performed a 
dynamical coupled-channels analysis of the available $p(e,e' \pi)N$ 
data in the region of $W \leq $ 1.6 GeV and $Q^2 \leq$ 1.45 (GeV/c)$^2$.
The channels included are $\gamma^* N$, $\pi N$, $\eta N$, and
$\pi\pi N$ which has  $\pi\Delta$, $\rho N$, and $\sigma N$
components. With the hadronic parameters of the model determined 
in our previous investigations of $\pi N\rightarrow \pi N$ reaction, 
we have found that the available data in the considered $W \leq $ 1.6 
GeV region can be fitted well by only adjusting the bare 
$\gamma^* N \rightarrow N^*$  helicity amplitudes for the lowest 
$N^*$ states in  $P_{33}$, $P_{11}$, $S_{11}$ and $D_{13}$ partial 
waves. The meson cloud effect, as required by the unitarity 
conditions, on the $\gamma^* N \rightarrow N^*$  form factors are 
examined.
\end{abstract}



\begin{multicols}{2}

\section{Introduction}

The electromagnetic parameters characterizing the excited nucleons, $N^*$, 
and in particular the $\gamma^* N \rightarrow N^*$ form factors, are important 
information for understanding the hadron structure within 
Quantum Chromodynamics (QCD). Thanks to the efforts in recent years, as 
reviewed in Ref.~\citep{bl04}, the world data of $\gamma^* N \rightarrow \Delta
(1232)$ form factors are considered along with the electromagnetic 
nucleon form factors as the benchmark data for developing hadron structure 
models and testing predictions from Lattice QCD calculations (LQCD). 

In the present work, we explore the extent to which the available 
$p(e,e'\pi)N$ data in $W\leq 1.6$~GeV can be used to extract the 
$\gamma^* N \rightarrow N^*$ form factors for the $N^*$ states up to the 
so-called ``second'' resonance region. 

We employ a dynamical coupled-channels model developed in 
Refs.~\citep{msl07,jlms07,djlss08,jlmss08,kjlms09} and extend our 
analysis~\cite{jlmss08} of pion photoproduction reactions.
We therefore will only recall equations which are relevant to 
the coupled-channels calculations of $p(e,e'\pi)N$ cross sections.
In the helicity-LSJ mixed-representation where the initial
$\gamma N$ state is specified by its helicities $\lambda_\gamma$ and
$\lambda_N$ and the final $MB$ states by the $(LS)J$ angular momentum
variables, the reaction amplitude of 
$\gamma^*(\vec{q}, Q^2) + N (-\vec{q}) \rightarrow
\pi(\vec{k}) + N (-\vec{k})$ at invariant mass $W$ and momentum transfer
$Q^2=-q^\mu q_\mu = \vec{q}^{\,2}-\omega^2 $ can be written within
a Hamiltonian formulation~\cite{msl07} as (suppress the isospin quantum numbers)
\begin{eqnarray}
T^{J}_{LS_N\pi N,\lambda_\gamma\lambda_N}(k,q,W,Q^2) =
{t}^{J}_{L S_N \pi N, \lambda_\gamma\lambda_N}(k, q, W,Q^2) &&\nonumber \\
+t^{R,J}_{LS_N\pi N,\lambda_\gamma\lambda_N}(k, q, W,Q^2)&&\,,
\label{eq:pw-t}
\end{eqnarray}
where $S_N=1/2$ is the nucleon spin, $W=\omega +E_N(q)$ is the invariant mass
of the $\gamma^* N$ system,  and the non-resonant amplitude is
\begin{eqnarray}
{t}^{J}_{L S_N \pi N,\lambda_\gamma\lambda_N}(k,q,W,Q^2)
=
{\it v}^{J}_{L S_N \pi N,\lambda_\gamma\lambda_N}(k,q,Q^2)&&\nonumber \\ 
+\sum_{M'B'}
\sum_{L^{\prime}S^{\prime}}
\int k^{\prime 2}dk^{\prime }
 {t}^{J}_{L S_N \pi N, L' S'M'B'}(k,k^{\prime},W) &&\nonumber \\
 \times G_{M'B'}(k',W)
{\it v}^{J}_{L' S' M'B' , \lambda_\gamma\lambda_N}(k',q,Q^2)&&\,.
\label{eq:pw-nonr}
\end{eqnarray}
In the above equation, $G_{M'B'}(k',W)$ are the meson-baryon propagators 
for the channels $M'B'= \pi N, \eta N, \pi\Delta, \rho N, \sigma N$. The 
matrix elements ${\it v}^{J}_{L S MB,\lambda_\gamma\lambda_N}(k,q,Q^2)$, 
which describe the $\gamma N \rightarrow MB$ transitions, are 
given explicitly in Appendix F of Ref.~\citep{msl07}. The hadronic 
non-resonant amplitudes ${t}^{J}_{L S_N \pi N, L' S'M'B'}(k,k^{\prime},W)$ 
are generated from the model constructed from analyzing the data of
$\pi N \rightarrow \pi N$ reactions~\cite{jlms07}.

The resonant amplitude in Eq.~(\ref{eq:pw-t}) is
\begin{eqnarray}
t^{R,J}_{LS_N\pi N,\lambda_\gamma\lambda_N}(k, q, W,Q^2) =
 \sum_{N^*_i, N^*_j}
[\bar{\Gamma}^{J}_{N^*_i,LS_N\pi N}(k,W)]^* &&\nonumber \\
D_{i,j}(W)
\bar{\Gamma}^{J}_{N^*_j,\lambda_\gamma\lambda_N}(q,W,Q^2) \,,&&
\label{eq:pw-r}
\end{eqnarray}
where the dressed $N^*\rightarrow \pi N$ vertex 
$\bar{\Gamma}^{J}_{N^*_i,LS_N\pi N}(k,W)$ and $N^*$ propagator $D_{i,j}(W)$
have been determined and given explicitly in Ref.~\citep{jlmss08}. The 
quantity relevant to our later discussions is the dressed
$\gamma ^* N \rightarrow N^*$ vertex function defined by
\begin{eqnarray}
\bar{\Gamma}^{J}_{N^*,\lambda_\gamma\lambda_N}(q,W,Q^2)
={\Gamma}^{J}_{N^*,\lambda_\gamma\lambda_N }(q,Q^2) &&\nonumber \\
+ \sum_{M'B'}
\sum_{L^{\prime}S^{\prime}}
\int k^{\prime 2}dk^{\prime }
 \bar{\Gamma}^{J}_{N^*,L'S'M'B'}(k',W) &&\nonumber \\
\times  G_{M'B'}(k',W)
{\it v}^{J}_{L' S' M'B' , \lambda_\gamma\lambda_N}(k',q,Q^2)\,.&& 
\label{eq:pw-v}
\end{eqnarray}

Similar to what was defined in Ref.~\citep{sl96,jlss07}, we call the 
contribution of the second term of Eq.~(\ref{eq:pw-v}) the 
{\it meson cloud effect} to define precisely what will be presented 
in this paper. We emphasize here that the meson cloud term in 
Eq.~(\ref{eq:pw-v}) is the necessary consequence of the unitarity
conditions. How this term and the assumed bare $N^*$ states
are interpreted is obviously model dependent.

Within the one-photon exchange approximation, the differential cross 
sections of pion electroproduction can be written as
\begin{eqnarray}
\frac{d\sigma^5}{d E_{e'} d\Omega_{e'} d\Omega_\pi^*}
=
\Gamma_\gamma
\left[ \sigma_T + \epsilon \sigma_L
 +\sqrt{2\epsilon(1+\epsilon)}\sigma_{LT} \cos \phi_\pi^\ast
\right.&&
\nonumber \\
\left.
+ \epsilon \sigma_{TT} \cos 2\phi_\pi^\ast
+ h_e\sqrt{2\epsilon(1-\epsilon)}
 \sigma_{LT^\prime}\sin \phi_\pi^\ast
\right].&&
\label{eq:dcrst-em}
\end{eqnarray}
Here $\Gamma_\gamma=[\alpha/(2\pi^2 Q^2)]
(E_{e'}/E_e)[|\vec{q}_L|/(1-\epsilon)]$;
$\epsilon$ is defined by the electron scattering angle $\theta_e$
and the photon 3-momentum ${\vec q}_L$ in the laboratory frame
as $\epsilon  =  [1 + 2 (|\vec{q}_L|^2/Q^2)\tan^2 (\theta_{e}/2)]^{-1}$;
$h_e$ is the helicity of the incoming electron;
$\phi_\pi^\ast$ is the angle between
the $\pi$-$N$ plane and the plane of the incoming and outgoing electrons.
The quantities associated with the electrons are defined
in the laboratory frame.
On the other hand, structure functions of $\gamma^* N\to\pi N$ process,
$\sigma_{\alpha}=\sigma_{\alpha}(W,Q^2,\cos\theta_\pi^\ast)$
($\alpha=T,L,LT,TT,LT')$, are defined in the final
$\pi N$ center of mass system.
The formula for calculating $\sigma_{\alpha}$
from the amplitudes defined by Eqs.~(\ref{eq:pw-t})-(\ref{eq:pw-r})
are given in Ref.~\citep{sl09}.

In this first-stage investigation, we only consider the data of
structure functions $\sigma_\alpha$ of $p(e,e'\pi^0)p$~\cite{m98,e1c-pi0ltp}
and $p(e,e'\pi^+)n$~\cite{e1c,e1c-pipltp}
up to $W=1.6$ GeV and $Q^2=1.45$ (GeV/c)$^2$.
The availability of the data in the corresponding $(W,Q^2)$ region
are found in Table~\ref{tab:data-list}. 
The resulting parameters are then confirmed against the original
five-fold differential cross section data~\cite{hallb-site}. 
This procedure could overestimate/underestimate the 
errors of our analysis, but is sufficient for the present 
exploratory investigation.

In section II, we present the results from our analysis.
Discussions on future developments are given in section III.

\begin{center}
\tabcaption{ \label{tab:data-list} $Q^2$ values, in GeV, for 
which there are available structure function data 
($Q^2\leq 1.45$ (GeV/c)$^2$). Data from 
Refs.~\protect\citep{m98,e1c-pi0ltp,e1c,e1c-pipltp}}
\footnotesize
\begin{tabular*}{80mm}{l@{\extracolsep{\fill}}|l|l}
                       & $\sigma_T+\epsilon\sigma_L,~\sigma_{LT},~\sigma_{TT}$
  &  $\sigma_{LT'}$\\
\hline
$\gamma^* p\to\pi^0 p$ & 0.4, 0.525, 0.65, 0.75, 0.9, 1.15, 1.46  & 0.4,0.65\\
$\gamma^* p\to\pi^+ n$ & 0.3, 0.4, 0.5, 0.6                 & 0.4,0.65\\
\bottomrule
\end{tabular*}
\end{center}

\end{multicols}
\ruleup

\begin{figure}[t]
\centering
\includegraphics[clip,width=12cm,angle=0]{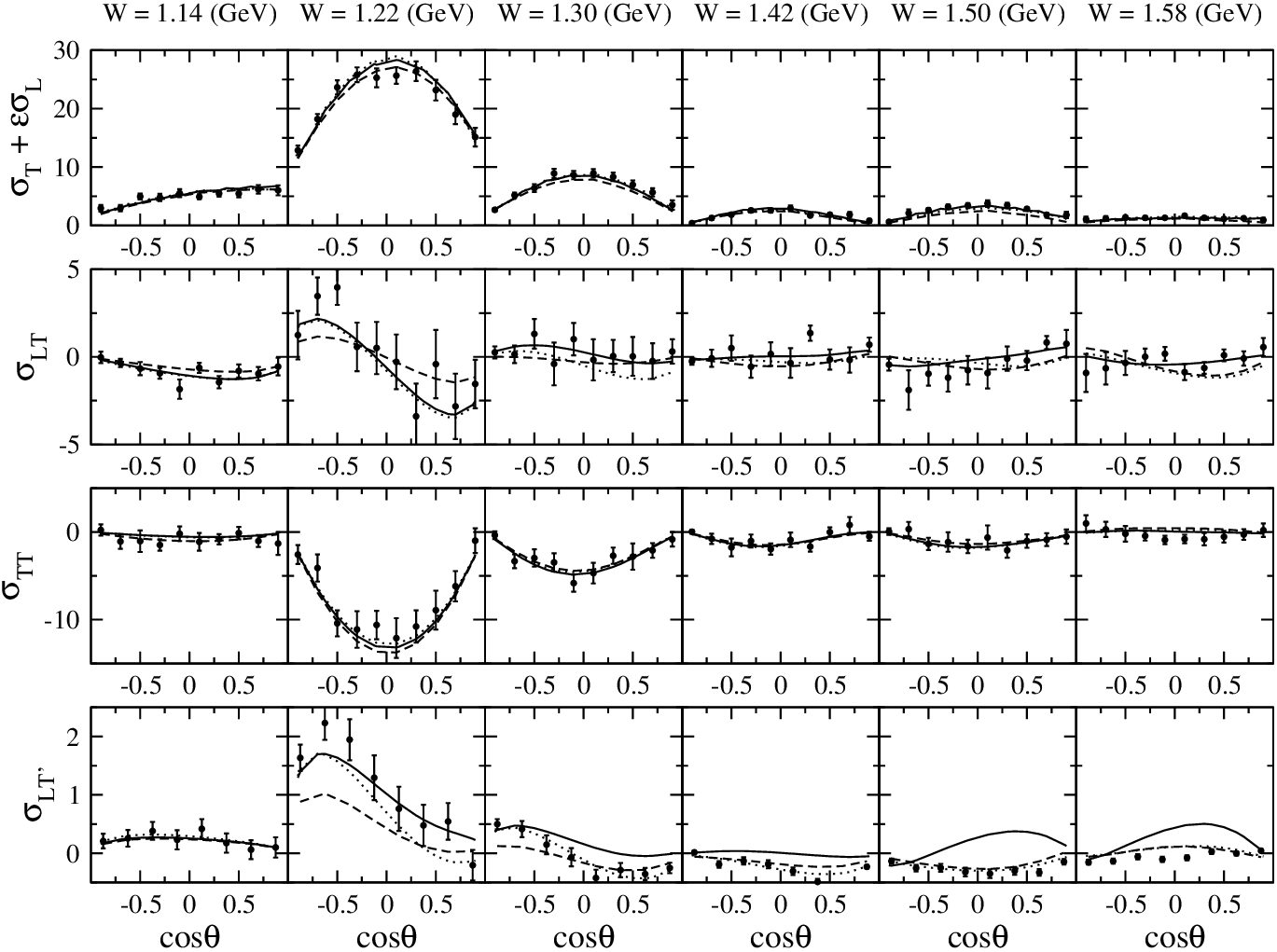}
\caption{Fit to $p(e,e'\pi^0)p$ structure functions at $Q^2=0.4$ (GeV/c)$^2$.
Here $\theta\equiv \theta_\pi^*$. The solid curves are the results of Fit1,
the dashed curves are of Fit2, and the dotted curves are of Fit3.
(See text for the description of each fit.)
The data are taken from Refs.~\protect\citep{m98,e1c-pi0ltp}.}
\label{fig:str-1}
\end{figure}

\ruledown
\begin{multicols}{2}

\section{Analysis and Results}

We parameterize the bare  $\gamma^\ast N \rightarrow N^*$ vertex
functions $\Gamma^{J}_{N^*,\lambda_\gamma\lambda_N}(q,Q^2)$ as
\begin{eqnarray}
{\Gamma}^{J}_{N^*,\lambda_\gamma\lambda_N}(q,Q^2)
=\frac{1}{(2\pi)^{3/2}}\sqrt{\frac{m_N}{E_N(q)}}\sqrt{\frac{q_R}{|q_0|}}
&&\nonumber \\
\times G_{\lambda}(N^*,Q^2) \delta_{\lambda, (\lambda_\gamma-\lambda_N)}, &&
\label{eq:ggn-a} 
\end{eqnarray}
where $q_R$ and $q_0$ are defined by 
$M_{N^*} = q_R+E_N(q_R)$ with $N^*$ mass and
$W = q_0+E_N(q_0)$, respectively, and 
\begin{eqnarray}
G_{\lambda}(N^*,Q^2) &=& A_{\lambda}(N^*,Q^2),~~~\text{transverse photon}, 
\label{eq:barea}\\
 &=& S_{\lambda}(N^*,Q^2),~~~\,\text{longitudinal photon}.
\label{eq:bares}
\end{eqnarray}
For later discussions, we also cast the helicity amplitudes of the
dressed vertex Eq.~(\ref{eq:pw-v}) into the form of Eq.~(\ref{eq:ggn-a}) 
with dressed helicity amplitudes
\begin{eqnarray}
\bar{A}_{\lambda}(N^*,Q^2) &=& {A}_{\lambda}(N^*,Q^2) + 
{A}^{\text{m.c.}}_{\lambda}(N^*,Q^2), \label{eq:dressa} \\
\bar{S}_{\lambda}(N^*,Q^2) &=& {S}_{\lambda}(N^*,Q^2) + 
{S}^{\text{m.c.}}_{\lambda}(N^*,Q^2), \label{eq:dresss}
\end{eqnarray}
where ${A}^{\text{m.c.}}_{\lambda}(N^*,Q^2)$ and
${S}^{\text{m.c.}}_{\lambda}(N^*,Q^2)$ are due to the 
meson cloud effect defined by the second term of Eq.~(\ref{eq:pw-v}).

With the hadronic parameters fixed in analyzing the $\pi N$ 
reaction data~\cite{jlms07,kjlms09}, the only freedom 
in analyzing the electromagnetic meson production reactions are 
the electromagnetic coupling parameters of the model. 
If the parameters listed in Ref.~\citep{msl07} are used to 
calculate the non-resonant interaction 
${v}^{J}_{L' S' M'B' , \lambda_\gamma\lambda_N}(k',q)$ in Eqs.~(\ref{eq:pw-nonr})
and~(\ref{eq:pw-v}), the only parameters to be determined from the 
data of pion electroproduction reactions are the bare helicity amplitudes
defined by Eq.~(\ref{eq:ggn-a}). Such a highly constrained analysis was 
performed in Ref.~\citep{jlmss08} for pion photoproduction. It was found
that the available data of $\gamma p \rightarrow \pi^0 p,~\pi^+ n$
can be fitted reasonably well up to invariant mass $ W\le 1.6$ GeV.
In this work we extend this effort to analyze the pion electroproduction
data in the same $W$ region.

We first try to fix the bare helicity amplitudes by fitting to
the data of $\sigma_T+\epsilon \sigma_L$, $\sigma_{LT}$, and
$\sigma_{TT}$ of $p(e,e'\pi^0)p$ in Ref.~\citep{m98} which covers 
almost all $(W,Q^2)$ region we are considering 
(see Table.~\ref{tab:data-list}). In a purely phenomenological 
approach, we first vary all of the helicity amplitudes of 16 
bare $N^*$ states, considered in analyzing the $\pi N \rightarrow \pi N,\pi \pi N$ 
data~\cite{jlms07,kjlms09}, in the fits to the data. It turns out 
that only the helicity amplitudes of the first $N^*$ states 
in $S_{11}$, $P_{11}$, $P_{33}$ and $D_{13}$ are relevant in 
the considered $W \leq$ 1.6 GeV. Thus only the bare 
helicity amplitudes associated with those four bare $N^*$ 
states (total 10 parameters) are varied in the fit and the 
other bare helicity amplitudes are set to zero. The numerical 
fit is performed at each $Q^2$ independently, using the MINUIT library.

The results of our fits are the solid curves in the top 
three rows of Fig.~\ref{fig:str-1}. Clearly our results 
from this fit agree with the data well. We obtain similar 
quality of fits to the data of Ref.~\citep{m98} at other 
$Q^2$ values listed in Table.~\ref{tab:data-list}. We have 
also used the magnetic $M1$ form factor of $\gamma^\ast N\to \Delta(1232)$ 
extracted from previous analyses as data for fitting.
We refer the results of this fit to as ``Fit1''.

\begin{figure}[t]
\centering
\includegraphics[clip,width=12cm,angle=0]{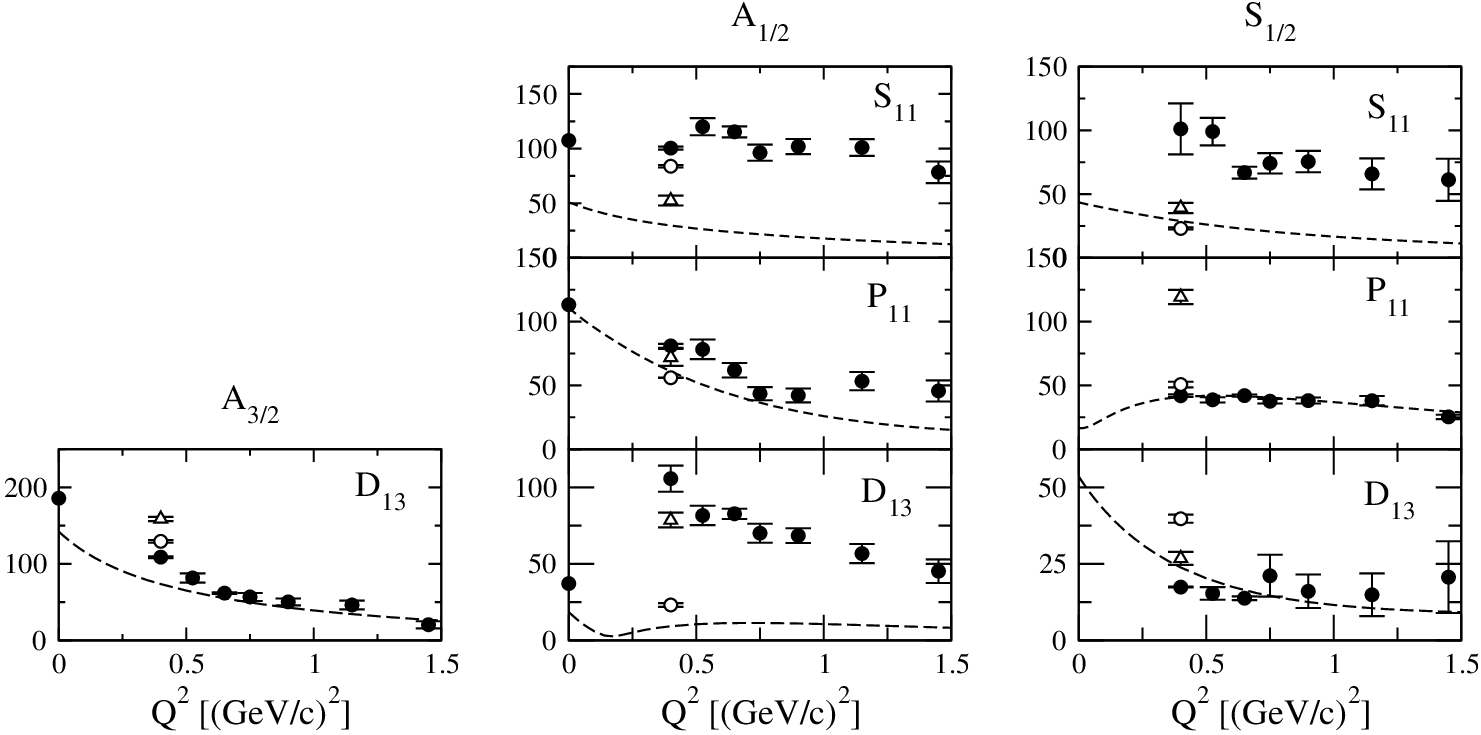}
\caption{Extracted helicity amplitudes for
$S_{11}$  at $W=1535$ MeV (upper panels),
$P_{11}$  at $W=1440$ MeV (middle panels),
and $D_{13}$  at $W=1520$ MeV (lower panels).
Solid points are from Fit1; dashed curves are the meson cloud contribution.
Open circles and triangles at $Q^2=0.4$ (GeV/c)$^2$
are from Fit2 and Fit3, respectively.
}
\label{fig:hel-s11p11d13}
\end{figure}

The helicity amplitudes of $S_{11}$, $P_{11}$, and $D_{13}$ resulting 
from Fit1 are shown in Fig.~\ref{fig:hel-s11p11d13}.
The solid circles are the absolute magnitude of the dressed helicity amplitudes
(\ref{eq:dressa}) and~(\ref{eq:dresss}).
The errors there are assigned by MIGRAD in the MINUIT library.
More detailed analysis of the errors is perhaps needed, but will not
be addressed here.
The meson cloud effect (dashed curves), as
defined by $A^{\text{m.c.}}_\lambda$ and $S^{\text{m.c.}}_\lambda$
of Eqs.~(\ref{eq:dressa}) and~(\ref{eq:dresss}) and
calculated from the second term of Eq.~(\ref{eq:pw-v}),
are the necessary consequence of 
the unitarity conditions. 
They do not include the bare helicity term determined here 
and are already fixed in the photoproduction analysis~\cite{jlmss08}.
Within our model (and within Fit1), 
the meson cloud contribution is relatively small
in $S_{11}$ and $A_{1/2}$ of $D_{13}$ even in the low $Q^2$ region.

Here we note that our helicity amplitudes defined in 
Eqs.~(\ref{eq:dressa}) and~(\ref{eq:dresss})
are different from the commonly used convention, say 
$A^{\text{cnv}}_\lambda$ and $S^{\text{cnv}}_\lambda$,
which are obtained from the imaginary part of the $\gamma^\ast N\to \pi N$
multipole amplitudes~\cite{abl08}.
This definition leads to helicity amplitudes which are real,
while our dressed amplitudes are complex.
It was shown in Ref.~\citep{sl01} that
for the $\Delta(1232)$ resonance our dressed helicity amplitudes 
(\ref{eq:dressa}) and (\ref{eq:dresss})
can be reduced to $A^{\text{cnv}}_\lambda $ and $S^{\text{cnv}}_\lambda $,
if we replace the Green function $G_{\pi N}$ with its principal value
in all loop integrals appearing in the calculation.
However, such reduction is not so trivial for higher resonance states
because the unstable $\pi\Delta,\rho N,\sigma N$ channels open,
and thus the direct comparison of the helicity amplitudes
from other analyses becomes unclear.

\begin{figure}[t]
\centering
\includegraphics[clip,width=12cm,angle=0]{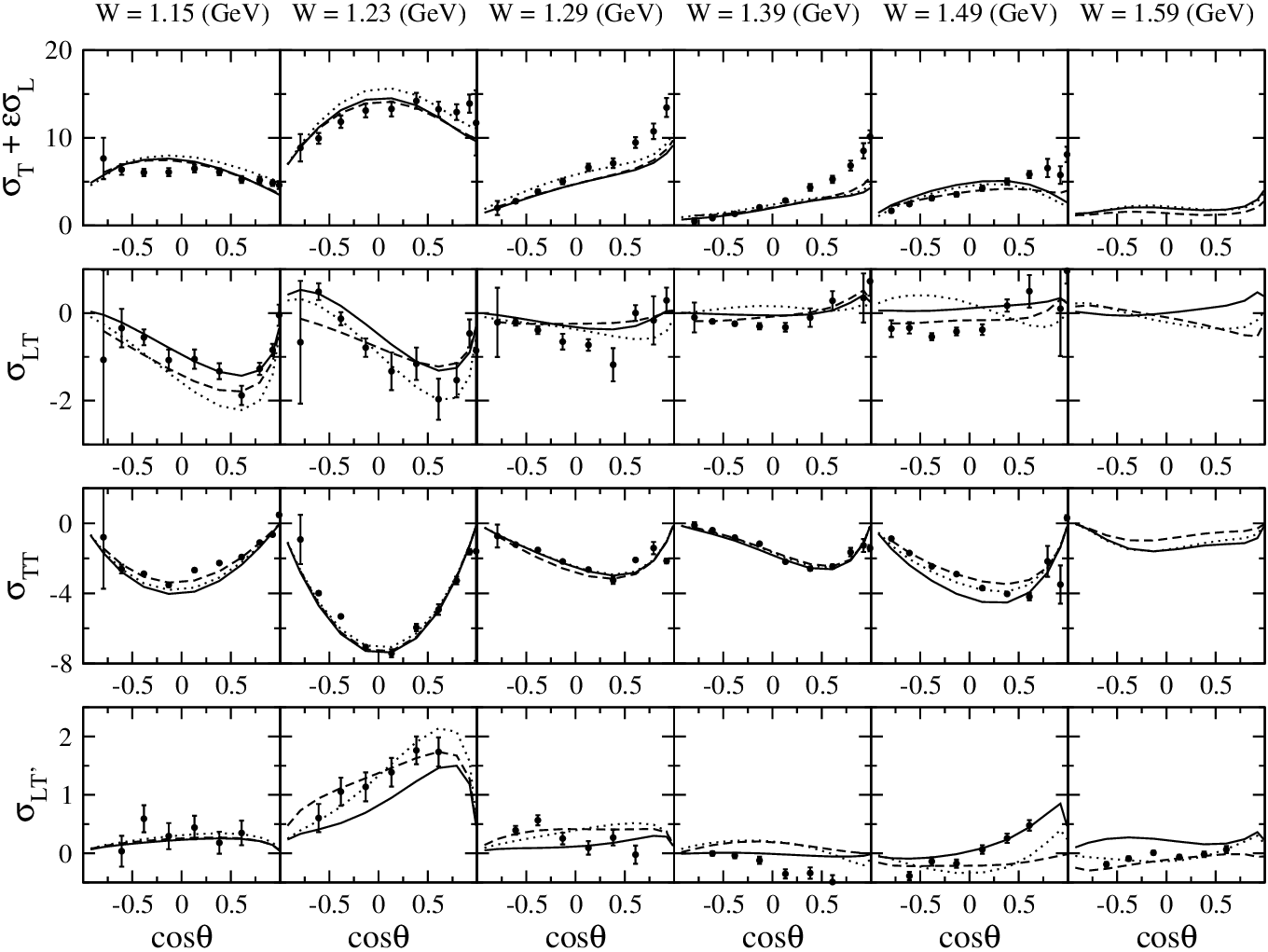}
\caption{
Structure functions of $p(e,e'\pi^+)n$ at $Q^2=0.4$ (GeV/c)$^2$.
Here $\theta\equiv\theta_\pi^*$.
The solid curves are the results of Fit1,
the dashed curves are of Fit2,
and the dotted curves are of Fit3.
(See text for the description of each fit.)
As for the $\sigma_{LT'}$, results at $W=1.14,1.22,1.3,1.38,1.5,1.58$ GeV
(from left to right of the bottom row) are shown, in which the data
are available.
The data in the figure are taken from Ref.~\protect\citep{e1c,e1c-pipltp}.
}
\label{fig:str-4}
\end{figure}

At $Q^2=0.4$ (GeV/c)$^2$, the data of all structure functions
both for $p(e,e'\pi^0)p$ and $p(e,e'\pi^+)n$ 
are available as seen in Table.~\ref{tab:data-list}. 
To see the sensitivity of the resulting helicity amplitudes to the
amount of the data included in the fits,
we further carry out two fits at this $Q^2$, 
referred to as Fit2 and Fit3, respectively.
Fit2 (Fit3) further includes the data of 
Refs.~\citep{e1c-pi0ltp,e1c,e1c-pipltp} (Ref.~\citep{e1c-pi0ltp})
in the fit in addition to those of Ref.~\citep{m98} which are used in Fit1.
This means that Fit2 includes all available data
both from $p(e,e'\pi^0)p$ and $p(e,e'\pi^+)n$, whereas
Fit3 includes the same data but from $p(e,e'\pi^0)p$ only.
The results of each fit are the dashed and dotted curves in 
Fig.~\ref{fig:str-1} for
$p(e,e'\pi^0)p$  and Fig.~\ref{fig:str-4} for $p(e,e'\pi^+)n$,
respectively. 

The corresponding change in the dressed helicity amplitudes are also shown
as open circles and triangles in Fig.~\ref{fig:hel-s11p11d13}. A significant 
change among the three different fits is observed in most of the results. 
This indicates that fitting 
the data listed in Table~\ref{tab:data-list} are far from
sufficient to pin down the $\gamma^* N \rightarrow N^*$ 
transition form factors up to $Q^2=1.45$ (GeV/c)$^2$. 
It clearly indicates the importance of
obtaining data from complete or over-complete measurements of most, if 
not all, of the independent $p(e,e'\pi)N$ 
polarization observables. 
Such measurements were made by Kelly
et al.~\cite{kelly} in the $\Delta$ (1232) region and
will be  performed at JLab for wide ranges of W and $Q^2$
in the next few years~\cite{bl04}.

\end{multicols}
\ruleup

\begin{figure}[t]
\centering
\includegraphics[clip,width=12cm,angle=0]{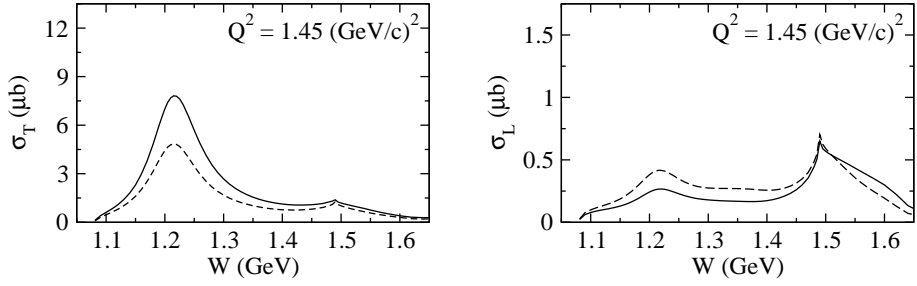}
\caption{
Coupled-channels effect on the integrated structure functions $\sigma_T(W)$
and $\sigma_L(W)$ for $Q^2=1.45 $ (GeV/c)$^2$ for $p(e,e'\pi^0)p$ reactions. 
The solid curves are the full results calculated with 
the bare helicity amplitudes of Fit1. The dashed curves are the same as 
solid curves but only the $\pi N$ loop is taken in the $M'B'$ summation 
in Eqs.~(\ref{eq:pw-nonr}) and~(\ref{eq:pw-v}).
 }
\label{fig:cc-pi0p}
\end{figure}

\ruledown
\begin{multicols}{2}

It has been seen in Fig.~\ref{fig:str-4} that all of our current fits
underestimate $\sigma_T$ of $p(e,e'\pi^+)n$ at forward angles.
We find that this can be improved by further varying 
the $S_{31}$ and $P_{13}$ bare helicity amplitudes 
within a reasonable range. We confirm that the same 
consequence is obtained also at other $W$,
and find that the $P_{13}$ ($S_{31}$) 
has contributions mainly at low (high) $W$.
We also find that the inclusion of the bare $S_{31}$ and $P_{13}$
helicity amplitudes does not change other structure functions than 
$\sigma_T$ of $p(e,e'\pi^+)n$ 
(the change is within the error).
This indicates that those two helicity amplitudes are rather relevant
to $p(e,e'\pi^+)n$, but not to $p(e,e'\pi^0)p$.
As shown in Table~\ref{tab:data-list}, however,
no enough data is currently available 
for $p(e,e'\pi^+)n$ above $Q^2=0.4$ (GeV/c)$^2$.
The data both of the 
$p(e,e'\pi^0)p$ and $p(e,e'\pi^+)n$ 
at same $Q^2$ values are desirable
to pin down the $Q^2$ dependence of the
$S_{31}$ and $P_{13}$ helicity amplitudes.

We now turn to show the coupled-channels effects (CCE). 
In Fig.~\ref{fig:cc-pi0p}, we see that when only the $\pi N$
intermediate state is kept in the $M'B'$ summation of
the non-resonant amplitude [Eq.~(\ref{eq:pw-nonr})] and
the dressed $\gamma^* N \to N^*$ vertices [Eq.~(\ref{eq:pw-v})], 
the predicted total transverse and longitudinal cross sections
$\sigma_T$ and $\sigma_L$ of $p(e,e'\pi^0)p$ are 
changed from the solid to dashed curves.
This corresponds to only examining
the CCE on the electromagnetic ($Q^2$-dependent) part
in the $\gamma^\ast N\to \pi N$ amplitude.
All CCE on the non-electromagnetic interactions
are kept in the calculations.
We find that the CCE 
tends to decrease when $Q^2$ increases.
This is rather clearly seen in $\sigma_T$.
In particular, the CCE on $\sigma_T$ at high 
$W\sim 1.5$ GeV is small ($10$-$20$\%) already at $Q^2 = 0.4 $ (GeV/c)$^2$.
(The effect is about $30$-$40$\% at $Q^2=0$.)
This is understood as follows.
In Eq.~(\ref{eq:pw-r}) we can further split the resonant amplitude $t^R$ 
as $t^R = t^R_{\text{bare}} + t^R_{\text{m.c.}}$,
where $t^R_{\text{bare}}$ and $ t^R_{\text{m.c.}}$
are the same as $t^R$ but replacing 
$\bar\Gamma^J_{N^\ast,\lambda_\gamma\lambda_N}$
with its bare part $\Gamma^J_{N^\ast,\lambda_\gamma\lambda_N}$
and meson cloud part [the second term of Eq.~(\ref{eq:pw-v})], respectively.
The CCE shown in Fig.~\ref{fig:cc-pi0p}
comes from $t^{J}_{LS_{N}\pi N,\lambda_\gamma\lambda_N}$ and $t^R_{\text{m.c.}}$.
We have found that the relative importance of 
the CCE
in each part remains the same for increasing $Q^2$.
However, the contribution of non-resonant mechanisms both on
$t^{J}_{LS_{N}\pi N,\lambda_\gamma\lambda_N}$ and $t^R_{\text{m.c.}}$
to the structure functions decreases for higher $Q^2$
compared with $t^R_{\text{bare}}$.
This explains the smaller CCE
compared with the photoproduction reactions~\cite{jlmss08}.
The decreasing non-resonant interaction at higher $Q^2$ is due to 
its long range nature, thus indicating that
higher $Q^2$ reactions provide a clearer probe of $N^\ast$.
We obtain similar results also for $p(e,e'\pi^+)n$.

It is noted, however, that the above argument does not mean 
CCE is negligible in the full $\gamma^\ast N\to \pi N$
reaction process.
In the above analysis we kept the CCE
on the hadronic non-resonant amplitudes, the strong $N^\ast$ vertices, 
and the $N^\ast$ self-energy, which are $Q^2$-independent
and remain important irrespective of $Q^2$.
We have found in the previous analyses~\cite{jlms07,kjlms09} 
that the CCE on them is significant in all energy region
up to $W= 2$ GeV.

\section{Summary and outlook}

We have explored how the available $p(e,e^\prime \pi)N$ data
can be used to determine the $\gamma^* N \rightarrow N^*$ transition
form factors within a dynamical coupled-channels 
model. Within the available data, the $\gamma^* N \rightarrow N^*$ bare 
helicity amplitudes of the first $N^*$ states in  $S_{11}$, $P_{11}$, 
$P_{33}$ and $D_{13}$ can be determined in the considered energy region, 
$W \leq$ 1.6 GeV. We further observe that some of these parameters can not
be determined  well. The uncertainties could be due to the limitation 
that only data of 4 out of 11 independent $p(e,e'\pi)N$ observables are 
available for our analysis. The data from the forthcoming measurements 
of double and triple polarization observables at JLab will be highly 
desirable to make progress.

We found that the underestimation of the $\sigma_T$ of $p(e,e'\pi^+)n$ 
at forward angles can be improved by further considering
the $S_{31}$ and $P_{13}$ bare helicity amplitudes. Furthermore, these 
amplitudes can have relevant contribution to $p(e,e'\pi^+)n$, but not 
to $p(e,e'\pi^0)p$. The $p(e,e'\pi^+)n$ data of wide $Q^2$ region
as well as $p(e,e'\pi^+)n$ seem necessary for determining the $Q^2$ 
dependence of the $S_{31}$ and $P_{13}$ helicity amplitudes.

For testing theoretical predictions from hadron structure calculations
such as LQCD, the quantities of interest are the residues of the
$\gamma^* N \rightarrow \pi N$ amplitudes, defined by 
Eqs.~(\ref{eq:pw-t})-(\ref{eq:pw-v}), at the corresponding resonance poles.
If the resonance poles are associated with the amplitude
$t^{R,J}_{LS_N\pi N,\lambda_\gamma\lambda_N}(k, q, W,Q^2)$ of
Eq.~(\ref{eq:pw-r}), the extracted residues are directly related to
the dressed form factors $\bar{\Gamma}^{J}_{N^*,L'S'M'B'}(k',W)$.
An analytic continuation method for extracting these information
has been developed~\cite{ssl09}, and our results along with
other hadronic properties associated to nucleon resonances
will be published elsewhere. Let us mention that 
the extracted form factors are complex numbers and
some investigations are needed to see how
they can be compared to the usual helicity amplitudes, which are real numbers,
listed by PDG~\cite{pdg}.

\acknowledgments{The author thanks H. Kamano, 
T.-S. H. Lee, A. Matsuyama and T. Sato, for the 
nice EBAC collaboration. This work is supported by a CPAN Consolider 
INGENIO CSD 2007-0042 contract and Grants No. FIS2008-1661 (Spain), by 
the U.S. Department of Energy, Office of Nuclear Physics Division, 
under contract No. DE-AC02-06CH11357, and Contract No. DE-AC05-06OR23177 
under which Jefferson Science Associates operates Jefferson Lab. This 
work used resources of the National Energy Research Scientific
Computing Center which is supported by the Office of Science of the 
U.S. Department of Energy under Contract No. DE-AC02-05CH11231.}

\end{multicols}

\vspace{-2mm}
\centerline{\rule{80mm}{0.1pt}}
\vspace{2mm}

\begin{multicols}{2}

\end{multicols}

\vspace{5mm}

\end{document}